\documentclass[a4paper,fleqn,usenatbib,useAMS]{mnras}
\usepackage{longtable}
\usepackage{mathptmx}

\usepackage[T1]{fontenc}
\usepackage{ae,aecompl}

\usepackage{etoolbox}
\makeatletter
\patchcmd\@combinedblfloats{\box\@outputbox}{\unvbox\@outputbox}{}{\errmessage{\noexpand patch failed}}
\makeatother
 
\usepackage{graphicx}	
\usepackage{amsmath}	
\usepackage{amssymb}	
\usepackage{multicol}   

\newcommand{\kms}{\,km\,s$^{-1}$} 
\newcommand{\meth}{CH$_{3}$OH} 
\newcommand{\target}{G358.93$-$0.03} 
\newcommand{\ngci}{NGC~6334I} 
\newcommand{\lsun}{L$_\odot$}

\makeatletter\ifdefined\Hy@StartlinkName
\def\addOneNestingLevelStartLink{%
  \gdef\Hy@StartlinkName##1##2{%
    \sbox0{\Hy@StartlinkNameOrig{##1}{##2}}\usebox0
    \global\let\Hy@StartlinkName\Hy@StartlinkNameOrig%
  }%
}
\def\addOneNestingLevelEndLink{%
  \gdef\pdfendlink{%
    \sbox0{\pdfendlinkOrig}\usebox0%
    \global\let\pdfendlink\pdfendlinkOrig%
  }%
}
\let\Hy@StartlinkNameOrig\Hy@StartlinkName
\let\pdfendlinkOrig\pdfendlink
\else
\let\addOneNestingLevelStartLink\relax
\let\addOneNestingLevelEndLink\relax
\fi

\title[New methanol maser transitions in \target]{Detection of new methanol maser transitions associated with \target}

\author[G. C. MacLeod et al.]
{
G. C. MacLeod$^{1, 2}$\thanks{E-mail: gord@hartrao.ac.za}, K. Sugiyama$^{3}$, T. R. Hunter$^{4}$,
J. Quick$^{2}$, W. Baan$^{5}$,    
\newauthor{
S. L. Breen$^{6}$, C. L. Brogan$^{4}$, 
R. A. Burns$^{3,7}$, A. Caratti o Garatti$^{8}$,
X. Chen$^{9,10}$, }
\newauthor{
J. O. Chibueze$^{11,12}$, M. Houde$^{1}$, J. F. Kaczmarek$^{13}$,
H. Linz$^{14}$, F. Rajabi$^{15,16}$, }
\newauthor{
Y. Saito$^{17}$, S. Schmidl$^{18}$, A. M. Sobolev$^{19}$,
B. Stecklum$^{18}$, }
\newauthor{
S. P. van den Heever$^{2}$, and Y. Yonekura$^{17}$ }\\
$^{1}$The University of Western Ontario, 1151 Richmond Street. London, ON N6A 3K7, Canada \\
$^{2}$Hartebeesthoek Radio Astronomy Observatory, PO Box 443, Krugersdorp, 1741, 
South Africa \\
$^{3}$Mizusawa VLBI Observatory, National Astronomical Observatory of Japan (NAOJ), 2-21-1 Osawa, Mitaka, Tokyo 181-8588,\\ 
Japan\\
$^{4}$NRAO, 520 Edgemont Rd, Charlottesville, VA, 22903, USA \\
$^{5}$Netherlands Institute for Radio Astronomy, ASTRON, 7991 PD, Dwingeloo, The Netherlands\\
$^{6}$Sydney Institute for Astronomy (SIfA), School of Physics, University of Sydney, NSW 2006, Australia\\
$^{7}$ Korea Astronomy and Space Science Institute, 776 Daedeokdae-ro, Yuseong-gu, Daejeon 34055, Republic of Korea\\
$^{8}$Dublin Institute for Advanced Studies, Astronomy \& Astrophysics Section, 31 Fitzwilliam Place, D02 XF86, Dublin 2, Ireland\\
$^{9}$Center for Astrophysics, GuangZhou University, Guangzhou 510006, China\\
$^{10}$Shanghai Astronomical Observatory, Chinese Academy of Sciences, 80 Nandan Road, Shanghai, 200030, China\\
$^{11}$Space Research Unit, Physics Department, North West University, Potchefstroom 2520, South Africa\\
$^{12}$Department of Physics and Astronomy, Faculty of Physical Sciences, University of Nigeria, Carver Building,\\ 1 University Road, 410001, Nsukka, Nigeria\\
$^{13}$CSIRO Astronomy and Space Science, CSIRO Parkes Observatory, PO Box 276 Parkes NSW
2870, Australia\\
$^{14}$Max Planck Institute for Astronomy, K\"onigstuhl 17, 69117 Heidelberg, 
Germany\\
$^{15}$Institute for Quantum Computing and Department of Physics and Astronomy, The
University of Waterloo, \\
200 University Ave. West, Waterloo, Ontario N2L 3G1, Canada\\
$^{16}$Perimeter institute for theoretical physics, Waterloo, ON, N2L 2Y5, Canada\\
$^{17}$Center for Astronomy, Ibaraki University, 2-1-1 Bunkyo, Mito, Ibaraki 310-8512, Japan\\
$^{18}$Th\"uringer Landessternwarte, Sternwarte 5, 07778 Tautenburg, Germany \\
$^{19}$Astronomical Observatory, Institute for Natural Sciences and Mathematics, Ural Federal University, 19 Mira street, Ekaterinburg,\\
620002, Russia\\
}
\date{Accepted 2019 August 22. Received 2019 August 20; in original form 2019 June 23}

\pubyear{2019}

\begin{document}
\label{firstpage}
\pagerange{\pageref{firstpage}--\pageref{lastpage}}
\maketitle

\begin{abstract}
We report the detection of new 12.178, 12.229, 20.347, and 23.121\,GHz methanol masers in the massive star-forming region \target, which are flaring on similarly short timescales (days) as the 6.668\,GHz methanol masers also associated with this source. The brightest 12.178\,GHz channel increased by a factor of over 700 in just 50 days. The masers found in the 12.229 and 20.347\,GHz methanol transitions are the first ever reported and this is only the fourth object to exhibit associated 23.121\,GHz methanol masers. The 12.178\,GHz methanol maser emission appears to have a higher flux density than that of the 6.668\,GHz emission, which is unusual. No associated near infrared flare counterpart was found, suggesting that the energy source of the flare is deeply embedded.

\end{abstract}

\section{Introduction}
Massive star formation is likely to involve episodic, disk-mediated bursts of accretion analogous to the FU Ori \citep{Hartmann96} and EX Ori \citep{Herbig77,Herbig89} phenomena seen in low mass stars.  The outbursts in these objects can occur over periods from weeks to decades \citep{Audard14}.
The deeply embedded nature of accreting massive protostars impedes observations and hampers direct investigation of the accretion process. However, recently, masers have emerged as a powerful tool to probe candidate accretion events. Two high-mass young stellar objects (HMYSO), \ngci-MM1 \citep{hetal17} and S255\,IR-NIRS3 \citep{Caratti_etal2017}, underwent major accretion events; the former found in the millimetric range and the latter in the infrared. In both cases 6.668\,GHz methanol maser flaring events were discovered serendipitously, but they occurred over time-scales of months \citep{macetal18, Fujisawa2015}.  At least in the case of \ngci, several other transitions with detected masers, including from other species, were monitored and also flared \citep{macetal18}.  

As a result of a single-dish maser monitoring program, \citet{Sugiyama19} reported a fast flaring event, rising on a timescale of days, in the 6.668\,GHz methanol masers associated with the massive star-forming region \target. But by all measures, \target\/ was a relatively unknown and unimpressive massive star-forming region. It was discovered via its associated 6.668\,GHz methanol masers by \citet{cetal2010}. They detected a maser with S$_{6.668}$(peak)=10\,Jy at the Local Standard of Rest velocity $v_{\rm LSR} = -15.9$\kms, after 2006 January 22 and before 2006 March 31; the velocity range of the weak emission was between $-22.0$ to $-14.5$\kms. A 12.178\,GHz methanol maser survey by \citet{breen2012} found no such associated emission  above 0.8\,Jy. No hydroxyl masers,  e.g. see \citet{Qiao14}, are reported. Weak water masers were detected by \citet{Titmarsh2016} where S$_{\rm 22\,GHz} \sim$ 0.7\,Jy at $v_{\rm LSR} = -21.6$\kms. \citet{Hu2016} mapped the 6.668\,GHz methanol masers of the high mass young stellar object (HMYSO) in 2012; it was only 5\,Jy and had the same velocity extent as originally reported. 
The masers were also mapped by \citet{Rickert2016} on 2015 September 09, who identified three 6.668\,GHz methanol maser features at $v_{\rm LSR} = -$18, $-$17, and $-$16\kms\/ ($F_{6.668}$ = 1 $-$ 3\,Jy) in a single position. 

Millmetre continuum emission from the HMYSO was reported in the Bolocam Galactic Plane Survey (BGPS) as source G358.936$-$0.032 \citep{Rosolowsky2010} and in the APEX Telescope Large Area Survey of the Galaxy (ATLASGAL) as source J174311.2$-$295129 \citep{Contreras2013}. \citet{Urquhart2013} reported a flux density of 1.4\,Jy at 870\,$\mu$m. 
The best estimate of the kinematic distance and bolometric luminosity range of the region is described by \citet{Brogan19} as $6.75^{+0.37}_{-0.68}$~kpc and 5700-22000~\lsun.  These authors present recent Atacama Large Millimeter Telescope (ALMA) and Submillimeter Array (SMA) data which reveal two hot molecular cores (MM1 and MM3 separated by about 1\farcs25), with the richer core exhibiting unprecedented (sub)millimetre methanol maser emission including torsionally excited (v$_t$ = 1 and 2) methanol transitions.  They state the brightness temperatures are high implying these are masers and not thermal emission lines. The thermal methanol emission toward this core peaks at $-16.5$\kms\/ with a linewidth of 3.1\kms.

The Maser Monitoring Organisation (M2O), including its voluntary group of observatories who monitor masers, was alerted to the rapidly strengthening 6.668\,GHz methanol masers associated with \target\/ \citep{Sugiyama19}. \citet{Breen19} amazingly discovered several never-before detected methanol maser transitions including torsionally excited (v$_{t}=1$) methanol transitions. 
We present confirmatory 6.668\,GHz methanol maser observations and the results of searches for hydroxyl, formaldehyde, water, and methanol masers associated with \target\/ using the Hartebeesthoek Radio Astronomy Observatory \mbox{(HartRAO)}. We also report the discovery of rare 23.121\,GHz methanol masers along with never-before detected masers in two other methanol transitions.  

\section{Observations}

\subsection{Hartebeesthoek Radio Astronomy Observatory}
Observations using the 26\,m telescope of Hartebeesthoek Radio Astronomy Observatory (HartRAO)\footnote{See http://www.hartrao.ac.za/spectra/ for further information.} were made in four receiver bands.  The 1.3, 4.5, 5.0, and 18.0\,cm receivers are each dual, left (LCP) and right (RCP), circularly polarised, cryogenically cooled receivers; the 2.5\,cm receiver was uncooled. Each receiver and polarisation were calibrated independently relative to Hydra~A and 3C123, assuming the flux scale of \citet{Ott94} (Jupiter was also observed for the 1.3\,cm receiver data). The point source sensitivity (PSS) for the 1.3 and 2.5\,cm receivers are 10.5 and 5.8\,Jy\,K$^{-1}$ per polarisation; for the 4.5, 5.0, and 18.0\,cm receivers it is 5.1\,Jy\,K$^{-1}$ per polarisation. 
The beamsize of the 1.3, 2.5, 4.5, 5.0, and 18.0\,cm receivers are 2\farcm1, 4\farcs0, 7\farcm0, 7\farcm5, and 29\farcm6 respectively. For all receivers, except the 1.3\,cm receiver, observations were completed in frequency switching mode and a 1.0\,MHz bandwidth on the 1024-channel (per polarisation) spectrometer. Observations made with the 1.3\,cm receiver employed position switching and a 2.0\,MHz bandwidth. Also half-power beamwidth pointing correction observations were completed for all epochs of observation except for the 18.0\,cm observations. More information for each receiver and the observing method employed are described in \citet{macetal18}.

The observations performed on 2019 January 21 included an attempt to confirm reports of the 6.668\,GHz methanol flare \citep{Sugiyama19}, and an exploratory investigation of the 12.178 and 23.121\,GHz lines. The velocity resolution of the 6.668, 12.178, and 23.121\,GHz observations are 0.044, 0.048, and 0.101\kms, respectively.  Monitoring began on 2019 January 25 and at a cadence of one to three days thereafter. Each monitoring epoch of observation is comprised of two six-minute observations. 

Prompted by our experience with \ngci\/ \citep{macetal18}, we also searched for associated hydroxyl (1.665, 1.667, 1.720, 1.612, 4.660, 4.765, 6.031, 6.035, and 6.049\,GHz), formaldehyde (4.830\,GHz) and water (22.235\,GHz) masers. The 1.665\,GHz OH transition was first observed on 2019 January 20 and at another seven epochs with the last on 2019 May 08; the others were observed between 2019 February 16 and March 25. The 4.830\,GHz formaldehyde transition was observed only once on 2019 February 17. Hydroxyl and formaldehyde observations were completed as described in \citet{macetal18}. 

Encouraged by the detection of new 12.178\,GHz masers and very rare 23.121\,GHz masers, we systematically searched for other new methanol transitions (83) in the available receivers described above beginning on 2019 March 12. Two new transitions with detectable masers, 12.229 and 20.347\,GHz were discovered on 2019 March 12 with a velocity resolution of 0.048 and 0.115\kms\/ respectively. For the 20.347\,GHz transition we observed using rest frequencies 20.346846\,GHz (provided by the Jet Propulsion Laboratory \citet{JPL1998}) and 20.346830\,GHz (F. J. Lovas, private communication and \citet{Remijan07}). We selected the latter because the spectra better matched the 6.668 and 12.178\,GHz methanol transition spectra. Monitoring of these transitions began shortly thereafter; results from this monitoring will be reported in a forthcoming paper. Transition information for each new maser is included in Table \ref{tab:g358_results}. Information of the non-detections (81) are listed in Table \ref{tab:g358_search}. The central frequencies observed were chosen predominantly from the list of transitions from tables made available by the Jet Propulsion Laboratory \citep{JPL1998}. Some were selected from two other sources: the Cologne Database for Molecular Spectroscopy \citep[CDMS;][]{CDMS2005} and from \citet{Lovas2004}. The uncertainty in the absolute flux density for all transitions was less than 6\%.

\subsection{Tianma Radio Telescope}

One service afforded by the M2O is to confirm new detections using different observatories. The 12.229\,GHz methanol masers were thus observed using the Shanghai 65\,m Tianma Radio Telescope (TMRT) on 2019 March 13; a day after the HartRAO discovery. A cryogenically cooled Ku-band receiver and digital backend system employing a 32768 channel spectral window yielding a velocity resolution of $\sim$0.018\kms\/ was used to obtain a typical rms noise of $\sim$0.1\,Jy per spectral channel. The beamsize was $\sim$1\farcm5. The uncertainty of the absolute flux density was less than 5\,\%. 

\subsection{Australia Telescope Compact Array}
Likewise, the Australia Telescope Compact Array (ATCA) observed this source on 2019 April 11 at the methanol line frequency of 20.3468300\,GHz. The ATCA was in the 750C array, where only antennas 1, 2, 3 and 4 were available, resulting in baseline lengths between 153 and 704\,m. The Compact Array Broadband Backend \citep[CABB;][]{Wilson-2011} was configured in CFB~1M$-$0.5k mode to provide 2\,MHz bandwidth with 4096 spectral channels, corresponding to a velocity coverage of 29\kms\/ and 0.007\kms\/ velocity channels.
Observations of G\,358.931$-$0.030 were interspersed with observations of nearby phase calibrator B\,1714$-$336 every 10 minutes, and the pointing was corrected every $\sim$50 minutes of observations. PKS~B\,1253$-$055 and PKS~B\,1934$-$638 were observed for bandpass and primary flux density calibration, respectively. The observations were conducted over a period of 6 hours, resulting in a total integration time of $\sim$2.5 hours on G\,358.931$-$0.030 and a synthesised beam of 13.0 $\times$ 2.14\,arcsec.
Data were reduced using {\sc miriad} \citep{miriad} following the procedure outlined in \citet{Breen19}. First a flux model was fit to the PKS\,B1253$-$055 continuum band data, which was then bootstrapped to PKS\,B1934$-$638 for absolute flux density calibration. The observations had sufficient parallactic angle coverage to use B\,1741$-$312 to calculate the leakage solution and we estimate that the polarisation calibration is accurate to within $\sim0.1$\,\% of Stokes I. The absolute flux density calibration is expected to be accurate to within 10\,\%.

\subsection{MPG/ESO 2.2m telescope}

Optical and near-infrared (NIR) imaging of \target\/ was performed employing the seven-channel Gamma-ray Burst Optical/Near-infrared Detector \citep[GROND;][]{Greiner2008}, using director's discretionary time (DDT) at the MPG/ESO 2.2\,m telescope at La Silla (Chile) on 2019 February 08 (Programme number 0103.C-9033(A)). GROND records imaging data in seven filters (optical: Sloan g' r' i' z' , near-infrared: J H Ks) simultaneously. The total integration time amounts to 38 minutes. Data processing was performed using the GROND pipeline \citep{Krüler2008}.

\section{Results}
\subsection{Non-detections}
We searched a number of transitions of OH (10), H$_2$CO (1), CH$_{3}^{~18}$OH (2), and CH$_3$OH (79) in which we found no emission. The 3$\sigma$ upper limits are presented in Table \ref{tab:g358_search}. The 1.665\,GHz OH transition was observed in eight epochs between 2019 January 20 and May 08 with no detections. The other 18\,cm OH transitions were re-observed in 2019 May, but we again report no detections.

\subsection{Detected methanol and water masers}

\begin{figure}   
	\includegraphics[width=\columnwidth]{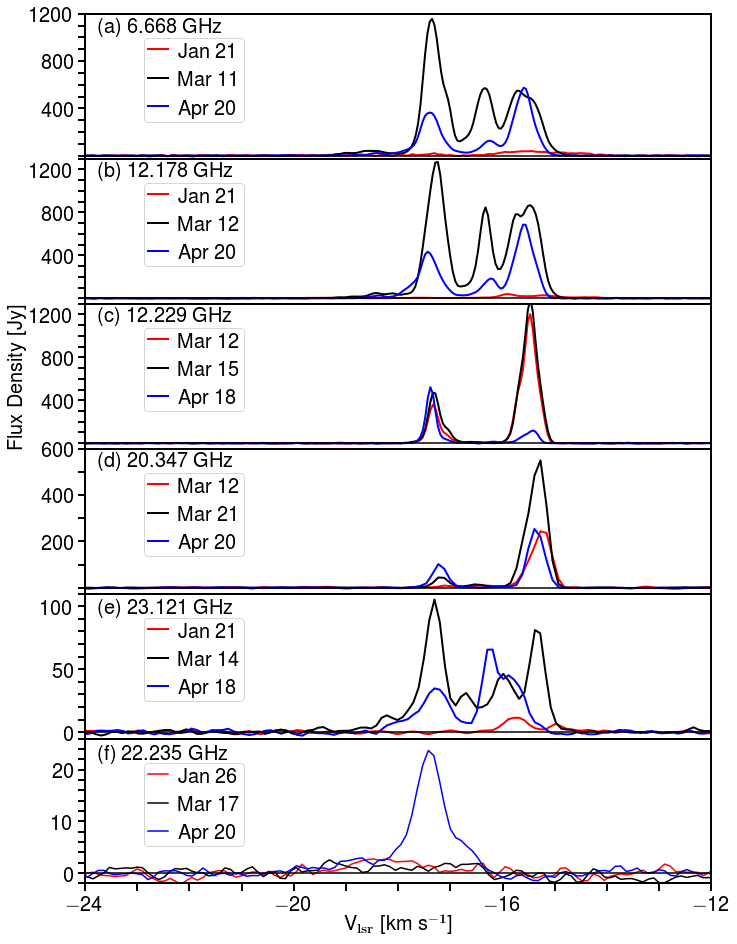}
	\caption{Selected spectra are plotted for: (a) 6.668\,GHz, (b) 12.178\,GHz, (c) 12.229\,GHz, (d) 20.237\,GHz, (e) 23.121\,GHz methanol, and (f) 22.235\,GHz water masers associated with \target.}
 \label{fig:g358_sp}
\end{figure}

We present spectra of all of the masers we detected associated with \target\/ in Fig.~\ref{fig:g358_sp}. Results of selected velocity channels of each transition are presented in Table~\ref{tab:g358_results}. We confirm the flaring of the 6.668\,GHz methanol maser emission and its fast variation reported by \citet{Sugiyama19} in the velocity range from $-$20.2 to $-$13.9\kms. This velocity extent is commensurate with the original detection of 6.668\,GHz methanol masers by \citet{cetal2010}. The 12.178 and 23.121\,GHz methanol spectra are similar to the 6.668\,GHz methanol maser spectra. \citet{Rickert2016} mapped the 6.668\,GHz emission and determined these spots were regions of maser activity; we assume here the similar 12.178 and 23.121\,GHz methanol emission also represent maser emission (more on this below). These 12.178 and 23.121\,GHz methanol masers are new detections towards this source. This is only the fourth object ever detected as a maser in the 23.121\,GHz transition, the others being W3(OH) \& NGC~7538 \citep{wwjs84} and \ngci\/ \citep{mb89}. For most of the velocity extent of the spectra we report that $F_{12.178} \ge F_{6.668} \ge F_{23.121}$, see the peaks listed in Table \ref{tab:g358_results}.  

Early results of our monitoring observations are included in Table \ref{tab:g358_results}; the flaring continues (MacLeod et al., in preparation). During these observations we find that the emission in the brightest 6.668\,GHz maser velocity channels increased by factors of between 17 and 66 in 71\,d and 49\,d respectively from 2019 January 21. The brightest velocity channel, $v_{\rm LSR} = -17.3$\kms, for the 12.178 and 23.121\,GHz methanol spectra each increased dramatically by factors of $\sim$700 and $\sim$200 in $\sim$50\,d.

The water maser emission appears within the velocity extent presented in \citet{Titmarsh2016}, and we initially detected it at a similar peak flux density level on 2019 January 26. It remained weak until flaring began between 2019 April 1 and 20; we confirm the onset of flaring reported by Xi et al. (in preparation). We estimate the water maser emission rose by a factor of $\sim$45.

In Fig.~\ref{fig:g358_sp} we also present the spectra of two new methanol maser transitions, 12.229 and 20.347\,GHz. These are the first such masers ever detected in these methanol transitions. The velocity extent of the 12.229 and 20.347\,GHz masers are both similar to that for the 6.668\,GHz masers. Their spectra are similar to each other but have only weak emission, $\leq$15\,Jy, in the velocity range between $-$16 to $-$16.8\kms\/ (unlike the other methanol transitions shown here). Our observations of these lines began at about the time the 6.668\,GHz methanol masers were peaking; the emission rose by factors between 1.1 and 7.0 (clearly lower limits). In summary, all three transitions are experiencing significant, contemporaneous, and fast flaring on timescales of days.

\begin{figure}   
	\includegraphics[width=\columnwidth]{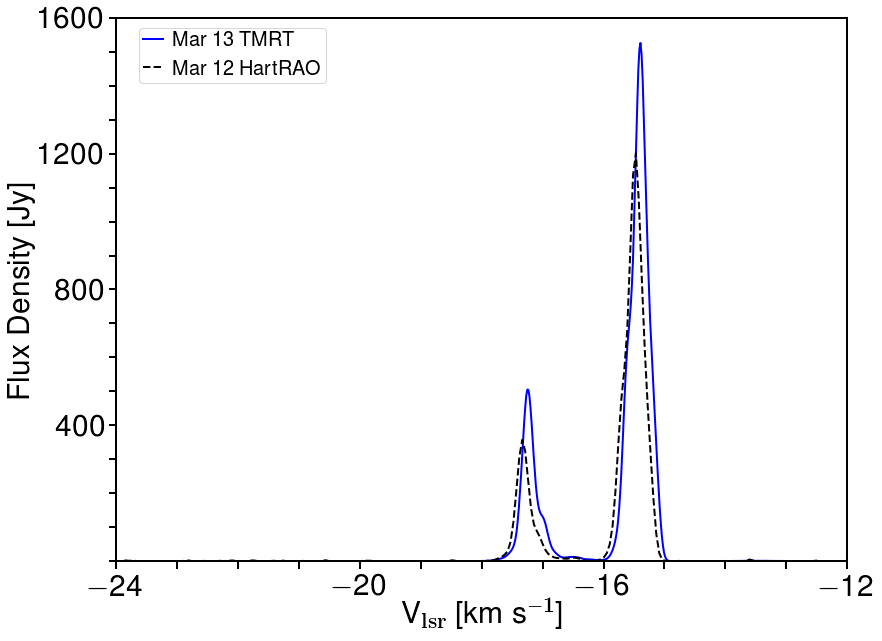}
	\caption{Spectra of 12.229\,GHz methanol masers associated with \target\/ are plotted for observations obtained by TMRT on 2019 March 13 (solid blue line) and by HartRAO, nearest the TMRT observation, on 2019 March 12 (dashed black line).}
 \label{fig:g358_Xi}
\end{figure}

\begin{figure}   
	\includegraphics[width=\columnwidth]{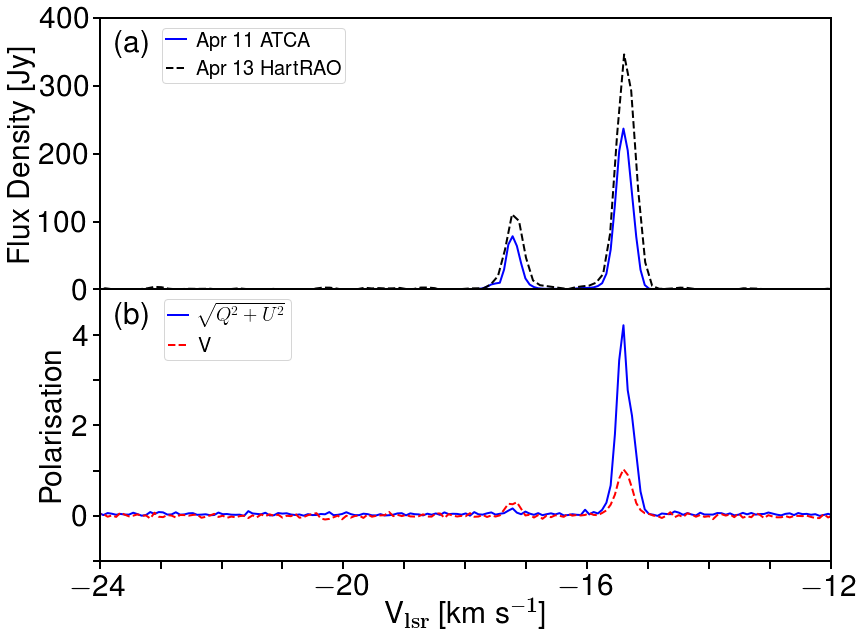}
	\caption{(a) Spectra of 20.347\,GHz methanol masers associated with \target\/ are plotted for observations obtained by the ATCA on 2019 April 11 (solid blue line) and by HartRAO, nearest the ATCA observation, on 2019 April 13 (dashed black line). (b) Linear (solid blue line and circular (dashed red line) polarisation data from the ATCA observations are plotted.}
 \label{fig:g358_Breen}
\end{figure}

Fortunately, observations of these new transitions were also made by others in the M2O group. We present the spectra from the TMRT (12.229\,GHz) in Fig.~\ref{fig:g358_Xi} and ATCA (20.347\,GHz) in Fig.~\ref{fig:g358_Breen}. The TMRT spectrum at 12.229\,GHz is similar but not identical to that obtained by HartRAO taken one day earlier. The brightest velocity channels at $v_{\rm LSR} = -17.3$ and $-$15.4\kms\/ in the TMRT observation are about 25 percent brighter than the interpolated values between the 2019 March 12 and 14 HartRAO observations. The difference may be the result of variability and/or differences in the calibration of each receiver. There is a slight velocity shift, $\sim 0.08$\kms, between the two spectra, possibly the result of slightly different transition frequencies used at each observatory. We estimate that the minimum brightness temperature for F$_{12.229}$(peak)~=~1500\,Jy is $\sim$1500\,K, more than three times the transition's lower energy level (451\,K). From the ATCA observations we find the temperature brightness is $\sim$26000\,K at 20.347\,GHz. Including the variability, similarity of the spectrum at different frequencies, and these estimates of the brightness temperature, we are certain the emission in each transition is maser emission.

At 20.347\,GHz we find that the two brightest velocity channels found in the HartRAO data are $\sim$1.5 times that seen in the ATCA spectrum. We made only moderate pointing and atmospheric corrections, about 15\,\% each, to the HartRAO spectrum shown; insufficient to account for the difference. Our ongoing monitoring observations suggest this transition peaked on 2019 March 21 and was weakening through 2019 April 13 thus suggesting the HartRAO flux density would have been greater on 2019 April 11 when the ATCA observations were taken. The ATCA observations, in the given configuration, would have included most of the extended emission. Barring significant calibration errors, it is possible 20.347\,GHz methanol masers weakened to 2019 April 11 and then rebrightened on 2019 April 13.  

The ATCA observations include polarisation information of the 20.347\,GHz methanol masers, we present this in Fig. \ref{fig:g358_Breen}. These masers are both circularly (Stokes parameter V) and linearly (Stokes parameters Q and U) polarised. Maximum polarisation is detected at $v_{\rm LSR} = -15.5$\kms\/ at $V \sim 1$\,\% and $\sqrt{Q^{2}+U^{2}} \sim 4$\,\%. At $v_{\rm LSR} = -17.2$\kms, the polarisation measures are less than one percent. This result is in line with the polarisation reported in \citet{Breen19} for several transitions including never-before-detected maser transitions. \citet{Breen19} also reported that each of the new transitions they detected, including 20.347\,GHz, were co-spatial with the 6.668\,GHz methanol masers and all are maser emission sources.

The maxima at $v_{LSR} \sim -17.3$\kms\/ in the 6.668, 12.178, and 23.121\,GHz transitions occurred on or about 2019 March 12 (MJD$_{ave} = 58554 \pm 1.5$). We find that for each transition the brightest red-shifted feature reached a maximum before the brightest blue-shifted feature (MJD values used are demarcated in boldface in Table \ref{tab:g358_results}). We determined the average of these estimated time lags is $\tau = 22\pm$8\,d (the uncertainty is the standard deviation).

More thorough analysis of our monitoring, and a better understanding of the evolution of this amazing flare, will be presented in a follow-up paper once the flare has subsided.

\subsection{Detection of a near-infrared point source}

GROND detected an NIR counterpart of the \target{} region in the J, H, and Ks bands, and not at shorter wavelengths. It is known as 2MASS~J17431001$-$2951460 \citep{Geballe19} and is also referred to as VVV~J174310.01$-$295146.12 in the Data Release 2 \citep{Minniti2017} of the VISTA Variable in the Via Lactea Survey (VVV). While 
2MASS did not detect the source in the H band, the VVV survey reports a detection at this wavelength, with H-Ks = 3.66 mag, i.e., a very red source. In fact, a lower resolution K band spectrum obtained with UKIRT at epoch 2008-08-18 \citep{Geballe19} shows a featureless spectrum with a continuum rising towards longer wavelengths, indicative of a dust-enshrouded environment. The absence of any spectral feature precludes conclusions on its spectral type. In the VVV survey, the object has shown rapid brightness changes with a 3$\sigma$ range of 0.4\,mag, and a peak-to-peak variation of 0.79\,mag during five years of VVV Ks-band monitoring. The GROND photometry indicated a Ks brightness elevated by 0.34\,mag with respect to the mean of 12.23\,mag, i.e., within the range of the normal variability. Its position, to within 0\farcs2, is consistent with the secondary hot core and dust continuum source MM3 detected by ALMA, which is located $\sim$1\farcs1 to the southwest of the main hot core MM1 \citep[cf.][]{Brogan19}.
Image subtraction of the latest VVV Ks frame (epoch 2015-09-19), after proper flux scaling and PSF matching, did not reveal any extended emission that would offer evidence of the existence of a light echo from an accretion outburst, unlike in the
case of S255IR-NIRS3 \citep{Caratti_etal2017}.


\begin{table*}
\caption{Information on the flaring maser transitions associated with \target. We determine, the peak, observation date of peak, the onset period, and increase factor in the selected velocity channels for each transition. We demarcate in boldface the MJD date where the flare reaches a maximum in each velocity channel we use to determine the average time lag.}
\label{tab:g358_results}
\begin{tabular}{cccccccccc}
\hline
\multicolumn{3}{c}{Transition} & \multicolumn{2}{c}{Velocity Extent} & \multicolumn{5}{c}{Velocity Channel Information}\\
\multicolumn{1}{c}{Molecule} & \multicolumn{1}{c}{Energy Level} & \multicolumn{1}{c}{Freq.} & \multicolumn{1}{c}{$v_{min}$} & \multicolumn{1}{c}{$v_{max}$} & 
\multicolumn{1}{c}{$v_{chan}$} & \multicolumn{1}{c}{$F_{peak}$} & \multicolumn{1}{c}{MJD$_{peak}$} &
\multicolumn{1}{c}{Onset} & \multicolumn{1}{c}{Increase} \\
 & & & & & & & & \multicolumn{1}{c}{Period} & \multicolumn{1}{c}{Factor} \\
  &  & \multicolumn{1}{c}{(GHz)} & \multicolumn{1}{c}{(\kms)} & \multicolumn{1}{c}{(\kms)} & 
 \multicolumn{1}{c}{(\kms)} & \multicolumn{1}{c}{(Jy)} & 
 \multicolumn{1}{c}{($+$50000)} & 
 \multicolumn{1}{c}{(days)} & \multicolumn{1}{c}{}\\
\hline

 \meth\/& $5_1 \longrightarrow 6_0~A^+$ (v$_t=$ 0)  & 6.668$^1$ & $-$20.2 & $-$13.9 & $-17.4$ & 1156 & \textbf{8553} & 49 & 66 \\
 & & &  &  & $-16.3$ & 615 & 8547 & 43 & 50 \\  
 & & &  &  & $-15.8$ & 828 & \textbf{8531} & 25 & 27 \\  
 & & &  &  & $-15.6$ & 589 & 8573 & 71 & 17\\  
  
 \meth\/  & $2_0 \longrightarrow 3_{-1}$~E (v$_t=$ 0) & 12.178$^1$ & $-20.5$ &$-13.6$ & $-17.2$ & 1270 & \textbf{8554} & 50 & 733.5\\
 & & &  & & $-16.3$ & 1256 & 8546 & 42 & 223.4 \\
 & & &  & & $-15.8$ & 1101 & \textbf{8530} & 26 & 35.0 \\
 & & &  & & $-15.6$ & 816 & 8546 & 42 & 43.6 \\

\meth\/ & $16_5 \longrightarrow 17_4$~E (v$_t=$ 0) & 12.229$^2$ & $-17.9$ & $-15.0$ & $-17.3$ & 1143 & \textbf{8572} & 16 & 3.2\\
 & & & & & $-15.5$ & 1344 & \textbf{8557} & 1 & 1.1\\
 
\meth\/ & $17_6 \longrightarrow 18_5$~E (v$_t=$ 0) & 20.347$^2$ & $-17.8$ & $-14.6$ & $-17.2$ & 136 & \textbf{8590} & 34 & 7.0\\
 & & & &  & $-15.3$ & 549 & \textbf{8563} & 7 & 1.1\\ 
 
\meth\/ & $9_2 \longrightarrow 10_1$~A$^+$ (v$_t=$ 0) & 23.121$^1$ & $-20.2$ & $-14.0$& $-17.3$ & 105 & \textbf{8556} & 51 & 207.2\\  
 & & & & &$-16.0$ & 60 & 8529 & 24 & 8.8\\ 
 & & & & &$-15.6$ & 73 & \textbf{8520} & 15 & 7.7\\ 
 & & & & &$-15.3$ & 102 & 8568 & 63 & 44.9\\ 
 
H$_2$O\/ & $6_1 \longrightarrow 5_2$~(F=5 $\longrightarrow$ 4, v=0)& 22.235$^1$ & $-20.2$ & $-14.0$& $-17.4$ & 42 & 8595 & $<$21 & $\sim$45\\  

\hline
\multicolumn{10}{l}{$^1$from the catalog of \citet{Lovas2004} and $^2$from the JPL Line Catalog  \citep{JPL1998}}\\
\end{tabular} 
\end{table*}

\subsection{Summary of results}
We confirm the fast flaring nature, varying on the day scale, of the 6.668\,GHz masers associated with the MMYSO \target. We also report the detection of new 12.178\,GHz methanol maser emission, only the fourth 23.121\,GHz methanol maser and that $F_{12.178} \ge$ $F_{6.668}$. Remarkably, we discovered new masers towards this source in the 12.229 and 20.347\,GHz methanol maser transitions. Observations at TMRT and ATCA confirm the existence of each, respectively. There appears to be a time lag, $\tau = 22\pm8$\,d, between when the brightest red-shifted (peaked first) and blue-shifted velocity channels. We report no maser emission in hydroxyl, formaldehyde and other methanol transitions. Lastly, we report little variation of the NIR emission associated with \target\/ and suggest that 2MASS~J17431001$-$2951460 is not the IR counterpart of the alleged bursting source. 

\section{Discussion}
\label{discussion}
\subsection{Temporal behaviour}
It is not clear if the maser emission in velocity channels listed in Table \ref{tab:g358_results} are co-propagating or co-spatial. However, there appears to be a time lag, $\tau = 22\pm$1\,d, between the two brightest velocity channels, $v=-17.4$ and $-15.8$\kms, in each of the transitions. At the near kinematic distance, $D_{kin} \sim 6.5$\,kpc, their maximum estimated separation is $\sim4000\pm200$\,au or 0\farcs6 assuming the flare is travelling at the speed of light and all transitions at the same velocity are within a few light days of each other. The projected separation of MM1 and 3 is 1\farcs1 \citep{Brogan19} while Brogan et al. and \citet{Breen19} find the masers are associated with MM1. It is also possible that the blue-shifted feature simply has a longer coherent path length and takes longer to reach its maximum. This argument is supported by the fact that $F_{peak}$($-$17.3) $>$ $F_{peak}$($-$15.8). More interferometric observations are required to aid our interpretation.

The 6.668\,GHz methanol maser results here confirm those reported by \citet{Sugiyama19}. In fact, the flare in each transition has risen quickly to each velocity channel maximum, in at most 70\,d from the start of observations, and reached flux density maxima of between $F_{23.121} = $60\,Jy  to $F_{12.229} = $1340\,Jy. The fact that they are all occurring in several transitions strongly supports the hypothesis that all are amplifying a common background exciting source and possibly co-propagating. The masers associated with the HMYSOs S255\,IR-NIRS3 \citep{Fujisawa2015} and \ngci\/ \citep{macetal18} also experienced strong flares and are likely caused by accretion events as reported by  \citet{hetal17} and  \citet{Caratti_etal2017} respectively. The former was identified via a major increase in the millimetric dust emission while the latter was found via its NIR brightening. No evidence for millimetric brightnening \citep{Brogan19} nor NIR brightening (this work) has been found during this flare in \target.

\subsection{The maser conditions}

Class II masers are radiatively pumped \citep[][and references therein]{sd94,csg05} and found in the vicinity of HMYSOs \citep[][and references therein]{Breen13}. An excellent theoretical description of the methanol molecule and its maser associated transitions is given in \citet{Sobolev97} and \citet{csg05}. They predict transitions in which maser emission can occur between the levels of the highly excited ground state and torsionally excited states of methanol. Searches for several of these predicted transitions were undertaken \citep[][and references therein]{Cragg01,Chipman16,Ellingsen11} which brought new detections but have shown that the masers in these transitions are quite rare. Here we find no maser emission, nor thermal emission, in the \citet{csg05} predicted methanol transitions at $\nu = $ 5.005, 20.171, and 20.909\,GHz.
Perhaps the flares in these transitions had already subsided before the respective observations  reported here in early 2019 March, or they did flare but at flux densities below the detection limits, $F_{\nu}$ $< 2 - 4$\,Jy. From the 3$\sigma$ rms values above, we estimate the corresponding brightness temperature upper limits in a hypothetical 1\arcsec\/ beam for $\nu = $ 5 and 20\,GHz to be $\sim10^5$\,K and $\sim10^4$\,K, respectively, easily sufficient for weaker masers to lurk undetected by the single dish. \citet{Breen19} found all the class II methanol masers they detected were co-spatial, within 0\farcs2, with an absolute positional accuracy of 0\farcs4 and associated with MM1 and the (sub)millimeter masers reported by \citet{Brogan19}.

\subsubsection{Known methanol transitions}
The conditions under which several class II methanol transitions are inverted, including the 6.668, 12.178, and 23.121\,GHz transitions discussed here, are presented in \citet[and references therein]{csg05}.  Models in \citet{csg05} predict that $F(12.178) > F(6.668)$ but only in a narrow range of conditions, i.e. the specific column density of methanol is high (N$_{meth}/\Delta V > 10^{14}$\,cm$^{-3}$\,s). In dynamic regions where the conditions are changing, we might not expect the conditions for the brighter 12.178\,GHz masers to exist for any significant length of time. This rarity seems confirmed by \citet{breen2012} who found that only about three percent of the 400 detected 6.668\,GHz masers had stronger 12.178\,GHz masers. They further suggest it is possible even these examples are only the result of variability and non-simultaneous observations. \citet{macetal18} did find a velocity extent in which $F(12.178) > F(6.668)$ in their near-simultaneous monitoring observations of \ngci. 

Here we report a clear example where the single-dish spectra of 12.178\,GHz is mostly brighter than its 6.668\,GHz counterpart observed within hours of each other. Likewise, the observations of the other transitions were taken within hours (or at most a day) of the 6.668\,GHz masers.  

\subsubsection{New methanol maser transitions}
\citet{Breen19} discovered six new methanol maser transitions including 6.181\,GHz, a transition not included in the comprehensive list of predicted methanol maser transitions in \citet{csg05}. Here we report two new methanol maser transitions, 12.229 and 20.347\,GHz while searching 86 transitions; neither are included in the list in \citet{csg05}. Not only are these never-before-seen methanol maser transitions but the former is brighter than all our other transitions reported here, S$_{12.229}$(peak) $>$ S$_{12.178}$(peak) $>$ S$_{6.668}$(peak). The latter is also strong, S$_{6.668}$(peak) $>$ S$_{20.347}$(peak) $>$ S$_{23.121}$(peak).  We detected no emission in ten hydroxyl transitions (nor the single formaldehyde transition we searched) in any epoch. 

A variety of physical conditions can produce transitions with detectable masers as described in \citet{csg02,csg05}. Since these new maser transitions have never been searched for before, we can only speculate on their rarity. It is not certain if these are even class II methanol maser transitions. Regardless, the conditions for methanol maser activity in \target\/ are possibly very unique. At the very least to predict a regime in which S$_{12.178}$(peak) $>$ S$_{6.668}$(peak) may also be one in which these other transitions can become inverted and generate maser emission. As noted above, this occurs when the specific column density of methanol is high \citep{csg05}. Such a condition might exist if these masers are located more deeply in the parental cloud where the gas and dust densities will be higher. The higher number density of hydrogen would explain why no hydroxyl masers were detected; \citet{csg02} propose that for number densities of hydrogen in the range 10$^{5}$ < n$_\mathrm{H}$ < 10$^{8.3}$\,cm$^{−3}$ hydroxyl and methanol masers are expected.   It is hoped that these data, and including the monitoring observations, will avail a better understanding of the variation of the masers as the flare evolves. We will expand upon this topic in our follow-up paper where our monitoring results are presented (MacLeod et al., in preparation).

\subsection{An accretion event?}
To date there are only two known accretion events that were accompanied by significant maser flaring and the discovery of rare masers. Direct evidence of an accretion event was first reported by \citet{Caratti_etal2017} in the HMYSO S255IR-NIRS3; they presented flaring in the near-infrared (NIR) and mid-infrared continuum, while \citet{Liu18} found submillimeter flaring. \citet{Fujisawa2015} found an associated 6.668\,GHz methanol maser flaring event in S255IR-NIRS3, which was characterized in detail by \citet{Szymczak18}. \citet{hetal17} proposed an accretion event as the progenitor in \ngci\/ from their detection of a sudden increase in the (sub)millimeter dust emission luminosity (a factor of $\sim70$). Again this event was accompanied by major flaring in ten maser transitions of three separate molecules, including some very rare, e.g. 4.660\,GHz OH and 23.121\,GHz methanol \citep{macetal18}. 

Thus far the masers in each transition associated with \target\/ appear to be varying contemporaneously at similar velocities, see Fig.~\ref{fig:g358_sp}. They have brightened significantly, by factors of hundreds and in days to weeks, and some are rare or never-before-detected masers. 
Dissimilar to the above two other HMYSOs, this source has no associated hydroxyl and only weak water maser emission. Perhaps the energy release during an accretion event is sufficient to energise the masers here and drive the rapid flaring, but the transitions with associated masers are a result of the conditions in which the coherent columns of gas reside. 

The idea that protostellar luminosity outbursts are associated with class II methanol masers is supported by modelling that suggests the masers are pumped by infrared radiation \citep{csg05}. However, we did not detect a NIR flare counterpart and nor did \citep{Brogan19} find evidence of an increase in the dust emission at millimeter wavelengths. Perhaps the exciting source which provides the pumping IR radiation is deeply embedded, similar to \ngci\/, and its flare had subsided before observations could be made. Above we find maser modelling requires higher hydrogen number densities to explain the masers detected; the relative strengths of masers reported here may support this argument. Recent simulations \citep{Meyer2019} indicate that during the early stages of HMYSO evolution, bursts often come in groups of successive events occurring sequentially. However, as HMYSO growth continues, the progression in envelope consumption/dispersal will eventually suppress the gravitational disk instability \citep{Vorobyov2015}. Thus, \ngci\/-type accretion bursts are expected to be more frequent than those resembling that of S255IR-NIRS3. If this amazing flaring event in \target\/ is the result of accretion onto the HMYSO, then it is the first to be detected solely by maser monitoring.

It is difficult to explain the cause of these very fast, strong methanol masers and at these new transitions. The possibility of
a rapid increase in the maser pumping efficiency that is accompanied
by only small changes in continuum emission should be explored akin to the arguments put forward in \citet{Sobolev97}.  Another possibility might be superradiance as described by \citet{Rajabi19} pertaining to the flaring methanol masers associated with S255IR-NIRS3. It is not clear what best describes the conditions of the maser environment nor this rapid variation presented here. 
Further insight will likely be gained
through future analysis of the
ongoing single dish light curves,
including a more accurate 
determination of time lags, along
with a detailed comparison of the
interferometric images of each
transition.

\section{Summary and Future Work}
We confirm that the 6.668\,GHz methanol masers associated with \target\/ are flaring on a timescale of days as was found in \citet{Sugiyama19}. We report the detection of new associated 12.178\,GHz methanol masers that are stronger than the associated 6.668\,GHz masers. These masers are accompanied by only the fourth 23.121\,GHz methanol maser and two never-before-detected methanol masers at 12.229 and 20.347\,GHz; this increases the number of rare methanol maser transitions detected towards \target. The brightest 12.178\,GHz channel increased by a factor of over 700 in just 50 days. All of the other transitions also experienced significant flaring activity that occurred on timescales of days to weeks similar to the associated 6.668\,GHz masers. We estimate that there is a time lag, $\tau = 22\pm8$\,d, between the two brightest velocity channels. Lastly, we detect no NIR flare counterpart. 

Revealing the cause for this extraordinary flaring activity is important for the interpretation of class II methanol flares in general. While it has been shown that these flares are signposts of accretion bursts, other excitation mechanisms might be at work as well. This enigmatic source will require interferometric observations and at many transitions, along with monitoring, to determine its nature.

\section*{Acknowledgments}
We thank Dr. Sugiyama for notifying the Maser Monitoring Organisation (M2O) about this exciting source and Dr. Jonathan Quick for his efforts to schedule time around various other observing programmes at HartRAO. I personally thank A\&D Stoneworks for the generous time off to work on this manuscript. This research has made use of the VizieR photometric viewer, CDS,
Strasbourg, France and the SAO/NASA Astrophysics Data System (ADS). We acknowledge the acceptance of our DDT time request by the director of MPIA. The National Radio Astronomy Observatory is a facility of the National Science Foundation operated under agreement by the Associated Universities, Inc. The Australia Telescope Compact Array is part of the Australia Telescope National
Facility which is funded by the Australian Government for operation as a National
Facility managed by CSIRO.

\bibliographystyle{mnras}
\bibliography{g35893_Detections} 

\appendix
\section{Frequency search list}
\label{Appendix A}
We searched transitions of OH (10), H$_2$CO (1), CH$_{3}^{~18}$OH (2), and CH$_3$OH (79) in which we found no emission. The results are presented in Table \ref{tab:g358_search}.

\clearpage
\onecolumn
\setlength{\LTcapwidth}{\linewidth} 
\begin{longtable}{cccccc}
\caption{List of transitions searched for associated masers with \target\/ but not detected.}
\label{tab:g358_search}
%
\\\hline
\multicolumn{2}{c}{Transition} & \multicolumn{2}{c}{Velocity Info.} & \multicolumn{1}{c}{Observation} &
\multicolumn{1}{c}{3$\sigma$ RMS}\\
\multicolumn{1}{c}{Freq.} &  \multicolumn{1}{c}{Energy Level} &    \multicolumn{1}{c}{V$_{coverage}$} &           \multicolumn{1}{c}{V$_{resolution}$} & \multicolumn{1}{c}{Date} & 
\multicolumn{1}{c}{} \\ 
\multicolumn{1}{c}{(GHz)} & & \multicolumn{1}{c}{(\kms)} & \multicolumn{1}{c}{(\kms)} & & \multicolumn{1}{c}{(Jy)} \\
\endfirsthead
\multicolumn{6}{c}%
{\tablename\ \thetable\ -- \textit{Continued from previous page}} \\
\hline
\multicolumn{2}{c}{} & \multicolumn{2}{c}{Velocity Info.} & \multicolumn{1}{c}{Observation} &
\multicolumn{1}{c}{Upper limit}\\
\multicolumn{1}{c}{Frequency} &  \multicolumn{1}{c}{Transition} &    \multicolumn{1}{c}{V$_{coverage}$} &           \multicolumn{1}{c}{V$_{resolution}$} & \multicolumn{1}{c}{Date} & 
\multicolumn{1}{c}{3$\sigma$ RMS} \\ 
\multicolumn{1}{c}{(GHz)} & & \multicolumn{1}{c}{(\kms)} & \multicolumn{1}{c}{(\kms)} & & \multicolumn{1}{c}{(Jy)} \\
\hline
\endhead
\hline \multicolumn{6}{r}{\textit{Continued on next page}} \\
\endfoot
\hline
\endlastfoot
\hline
\multicolumn{6}{l}{Hydroxyl (OH)}\\
1.61223101$^1$ & N=$1^{-}\longrightarrow1^{+}$, J=3/2$\longrightarrow$3/2, F=$1\longrightarrow2$ & 47 & 0.091 & 2019 Feb 16 & 2.0 \\
1.66540184$^1$ & N=$1^{-}\longrightarrow1^{+}$, J=3/2$\longrightarrow$3/2, F=$1\longrightarrow1$ & 45 & 0.088 & 2019 Feb 20 & 5.0 \\
1.66735903$^1$ & N=$1^{-}\longrightarrow1^{+}$, J=3/2$\longrightarrow$3/2, F=$2\longrightarrow2$ & 45 & 0.088 & 2019 Feb 17 & 4.0 \\
1.72052998$^1$ & N=$1^{-}\longrightarrow1^{+}$, J=3/2$\longrightarrow$3/2, F=$2\longrightarrow1$ & 44 & 0.085 & 2019 Feb 16 & 2.0 \\
4.660242$^1$ & N=$1^{+}\longrightarrow1^{-}$, J=1/2$\longrightarrow$1/2, F=$0\longrightarrow1$ & 32 & 0.063 & 2019 Mar 21 & 0.5 \\ 
4.765562$^1$ & N=$1^{-}\longrightarrow1^{+}$, J=1/2$\longrightarrow$1/2, F=$1\longrightarrow0$ & 32 & 0.061 & 2019 Mar 11 & 1.0 \\ 
6.016746$^1$ &  N=$2^{+}\longrightarrow2^{-}$, J=5/2$\longrightarrow$5/2, F=$2\longrightarrow3$ & 25 & 0.049 & 2019 Mar 25 & 1.0 \\
6.030747$^1$ & N=$2^{+}\longrightarrow2^{-}$, J=5/2$\longrightarrow$5/2, F=$2\longrightarrow2$ & 24 & 0.049 & 2019 Mar 11 & 4.0 \\ 
6.035092$^1$ & N=$2^{+}\longrightarrow2^{-}$, J=5/2$\longrightarrow$5/2, F=$3\longrightarrow3$ & 24 & 0.049 & 2019 Mar 11 & 4.0 \\
6.049084$^1$ & N=$2^{+}\longrightarrow2^{-}$, J=5/2$\longrightarrow$5/2, F=$3\longrightarrow2$ & 24 & 0.048 & 2019 Mar 27 & 1.5 \\ 
\hline
\multicolumn{6}{l}{Formaldehyde (H$_{2}$CO)}\\
4.829659$^2$ & 1(1,0) $\longrightarrow$ 1(1,1), F=$2 \longrightarrow 2$ & 30 & 0.061 & 2019 Feb 17 & 1.0 \\
\hline
\multicolumn{6}{l}{Methanol (CH$_{3}^{~18}$OH)}\\
1.617244$^2$ & $18_5 \longrightarrow 18_9$~A$^+$ (v$_t$= 1 $\rightarrow$ 0) & 45 & 0.088 & 2019 May 16 & 3.0 \\
1.654342$^2$ & $22_2 \longrightarrow 22_2$~A$^+$ (v$_t$=1) & 45 & 0.088 & 2019 May 17 & 3.0 \\
\hline
\multicolumn{6}{l}{Methanol (CH$_{3}$OH)}\\
4.797547$^3$ & $28_2$ $\longrightarrow$ $28_2$~A$^{-+}$, (v$_t$=1) & 32 & 0.061 & 2019 Mar 14 & 2.0 \\
4.912974$^3$ & $35_2$ $\longrightarrow$ $35_1$~E (v$_t$=0) & 34 & 0.060 & 2019 Mar 14 & 2.0 \\
4.9552$^3$ & $25_4$ $\longrightarrow$ $26_1$~A$^-$ (v$_t$=0) & 34 & 0.059 & 2019 Mar 14 & 2.0 \\
5.005302$^3$ & $3_1\longrightarrow3_1$~A$^{-+}$ (v$_t$=0) &  34 & 0.058 & 2019 Mar 12 & 2.0 \\
5.091373$^3$ & $7_1 \longrightarrow 7_1$~A$^{-+}$ (v$_t$=2) &  34 & 0.058 & 2019 Mar 14 & 2.0 \\
6.107493$^3$ & $35_4 \longrightarrow 36_1$~A$^{+}$ (v$_t$=0) & 34 & 0.048 & 2019 Mar 14 & 2.0 \\
6.27083$^3$ & $30_2 \longrightarrow 30_2$~A$^{-+}$ (v$_t$=1) & 34 & 0.047 & 2019 Mar 14 & 2.0 \\
6.328305$^3$ & $11_1 \longrightarrow 11_1$~A$^{-+}$ (v$_t$=1) & 34 & 0.046 & 2019 Mar 15 & 2.0 \\
6.335727$^3$ & $37_{15} \longrightarrow 36_{14}$~A$^{+-}$ (v$_t$=1 $\longrightarrow$ 2) & 34 & 0.046 & 2019 Mar 14 & 2.0 \\
6.429225$^3$ & $19_4 \longrightarrow 18_3$~A$^+$ (v$_t$=2) & 34 & 0.046 & 2019 Mar 14 & 2.0 \\
6.543733$^3$ & $8_1 \longrightarrow 8_1$~A$^{-+}$ (v$_t$=2) & 34 & 0.045 & 2019 Mar 14 & 2.0 \\
6.745123$^3$ & $25_3 \longrightarrow 25_3$~A$^{-+}$ (v$_t$=0) & 34 & 0.043 & 2019 Mar 14 & 3.0 \\
6.758772$^3$ & $36_4 \longrightarrow 36_4$~A$^{+-}$ (v$_t$=0) & 34 & 0.043 & 2019 Mar 15 & 3.0 \\
6.785303$^3$ & $28_5 \longrightarrow 28_9$~A$^+$ (v$_t$=1 $\longrightarrow$ 0) & 34 & 0.043 & 2019 Mar 15 & 3.0 \\
8.176405$^3$ & $9_1 \longrightarrow 9_1$~A$^{-+}$ (v$_t$=2) &  34 & 0.036 & 2019 Mar 14 & 2.0 \\
8.254164$^3$ & $37_4 \longrightarrow 37_4$~A$^{+-}$ (v$_t$=0) &  34 & 0.036 & 2019 Mar 14 & 2.0 \\
8.34159$^3$ & $4_1 \longrightarrow 4_1$~A$^{-+}$ (v$_t$=0) &  34 & 0.036 & 2019 Mar 12 & 1.0 \\
8.465756$^3$ & $26_3 \longrightarrow 26_3$~A$^{-+}$ (v$_t$=0) &  34 & 0.035 & 2019 Mar 14 & 2.0 \\
8.467421$^3$ & $6_2 \longrightarrow 5_{-1}$~E (v$_t$=1) &  34 & 0.035 & 2019 Mar 14 & 2.0 \\
8.511071$^3$ & $38_{102} \longrightarrow 37_{103}$~E (v$_t$=1) &  34 & 0.034 & 2019 Mar 14 & 2.0 \\
8.523464$^3$ & $29_4 \longrightarrow 28_1$~A$^{+}$ (v$_t$=1) &  34 & 0.034 & 2019 Mar 14 & 2.0 \\
8.655609$^3$ & $17_2 \longrightarrow 17_2$~A$^{+-}$ (v$_t$=0) &  34 & 0.034 & 2019 Mar 14 & 2.0 \\
8.710622$^3$ & $13_1 \longrightarrow 13_1$~A$^{-+}$ (v$_t$=1) &  34 & 0.034 & 2019 Mar 14 & 2.0 \\
8.724144$^3$ & $40_{13} \longrightarrow 39_{14}$~E (v$_t$=2) &  34 & 0.034 & 2019 Mar 14 & 2.0 \\
8.845018$^3$ & $15_5 \longrightarrow 14_7$~E (v$_t$=2) &  35 & 0.033 & 2019 Mar 14 & 2.0 \\
8.856425$^3$ & $26_4 \longrightarrow 27_5$~A$^{-}$ (v$_t$=1) &  34 & 0.033 & 2019 Mar 14 & 2.0 \\
8.859549$^3$ & $26_4 \longrightarrow 27_5$~A$^{+}$ (v$_t$=1) &  34 & 0.033 & 2019 Mar 14 & 2.0 \\
11.842485$^3$ & $32_2 \longrightarrow 32_1$~E (v$_t$=0) &  34 & 0.049 & 2019 Mar 14 & 2.0 \\
11.964007$^3$ & $33_{16} \longrightarrow 34_{13}$~E (v$_t$=0 $\longrightarrow$ 1) &  35 & 0.049 & 2019 Mar 14 & 4.0 \\
11.980899$^3$ & $11_1 \longrightarrow 11_1$~A$^{-+}$ (v$_t$=2) &  34 & 0.049 & 2019 Mar 14 & 3.0 \\
12.058717$^3$ & $39_4 \longrightarrow 39_4$~A$^{+-}$ (v$_t$=0) &  34 & 0.049 & 2019 Mar 14 & 3.0 \\
12.329348$^3$ & $16_5 \longrightarrow 17_4$~E (v$_t$=0) &  34 & 0.048 & 2019 Mar 14 & 2.0 \\
12.347739$^2$ & $25_5 \longrightarrow 24_6$~A$^+$ (v$_t$=0) &  34 & 0.047 & 2019 Mar 14 & 3.0 \\
12.351$^2$ & $25_5 \longrightarrow 24_6$~A$^{-}$ (v$_t$=0) &  34 & 0.047 & 2019 Mar 14 & 3.0 \\
19.9673961$^1$ & $2_1 \longrightarrow 3_0$~E (v$_t$=0) &  60 & 0.117 & 2019 Mar 08 & 4.0 \\
20.171205$^3$ & $11_1 \longrightarrow 10_2$~A$^+$ (v$_t$=0)  &  59 & 0.116 & 2019 Mar 04 & 4.0 \\
20.182358$^3$ & $22_5 \longrightarrow 22_9$~A$^+$ (v$_t$=1 $\longrightarrow$ 0) &  59 & 0.116 & 2019 Mar 17 & 3.0 \\
20.204441$^3$ & $39_{-2} \longrightarrow 40_{-5}$~E (v$_t$=2 $\longrightarrow$ 1) &  59 & 0.116 & 2019 Mar 17 & 2.0 \\
20.407849$^3$ & $11_3 \longrightarrow 10_1$~A$^+$ (v$_t$=2)  &  59 & 0.116 & 2019 Mar 16 & 4.0 \\
20.908817$^3$ & $16_{-4} \longrightarrow 15_{-5}$~E (v$_t$=0) &  57 & 0.112 & 2019 Mar 12 & 3.0 \\
21.146747$^3$ & $8_2 \longrightarrow 7_3$~E (v$_t$=1) &  56 & 0.111 & 2019 Mar 17 & 3.0 \\
21.169082$^3$ & $18_9 \longrightarrow 19_{10}$~E (v$_t$=2) &  56 & 0.111 & 2019 Mar 17 & 2.0 \\
21.184378$^3$ & $24_{-4} \longrightarrow 25_{-2}$~E (v$_t$=0) &  56 & 0.111 & 2019 Mar 16 & 3.0 \\
21.26472$^3$ & $15_3 \longrightarrow 16_1$~A$^{+}$ (v$_t$=0) &  56 & 0.110 & 2019 Mar 16 & 3.0 \\
21.281616$^3$ & $19_{16} \longrightarrow 18_{18}$~E (v$_t$=1 $\longrightarrow$ 0) &  56 & 0.110 & 2019 Mar 16 & 3.0 \\
21.295299$^3$ & $31_{15} \longrightarrow 30_{18}$~E (v$_t$=2 $\longrightarrow$ 1) &  56 & 0.110 & 2019 Mar 18 & 2.0 \\
21.326889$^3$ & $10_{-9} \longrightarrow 11_{-8}$~E (v$_t$=2) &  56 & 0.110 & 2019 Mar 18 & 3.0 \\
21.443345$^3$ & $13_6 \longrightarrow 12_3$~E (v$_t$=1 $\longrightarrow$ 2) &  56 & 0.109 & 2019 Mar 18 & 6.0 \\
21.692261$^3$ & $36_4 \longrightarrow 37_{-3}$~E  (v$_t$=1) &  55 & 0.109 & 2019 Mar 18 & 2.0 \\
21.727015$^{3}$ & $33_5 \longrightarrow 33_9$~A$^+$ (v$_t$=1 $\longrightarrow$ 0) &  55 & 0.108 & 2019 Mar 18 & 2.0 \\
21.844244$^{3}$ & $28_0 \longrightarrow 27_{-2}$~E (v$_t$=1) &  55 & 0.107 & 2019 Mar 18 & 3.0 \\
21.888175$^{3}$ & $21_1 \longrightarrow 21_2$~A$^{-+}$ (v$_t$=1) &  55 & 0.107 & 2019 Mar 18 & 2.0 \\
21.932058$^{3}$ & $19_2 \longrightarrow 18_4$~A$^-$ (v$_t$=0) &  54 & 0.107 & 2019 Mar 18 & 3.0 \\
22.019094$^{3}$ & $11_3 \longrightarrow 10_5$~A$^+$ (v$_t$=1) &  54 & 0.106 & 2019 Mar 18 & 3.0 \\
22.111687$^{3}$ & $25_{-5} \longrightarrow 26_{-3}$~E (v$_t$=0) &  54 & 0.106 & 2019 Mar 18 & 4.0 \\
22.200055$^{3}$ & $32_{17} \longrightarrow 31_{19}$~A$^{+-}$ (v$_t$=1 $\longrightarrow$ 0) &  54 & 0.106 & 2019 Mar 18 & 3.0 \\
22.3012$^{3}$ & $31_{101} \longrightarrow 30_{103}$~E (v$_t$=2) &  54 & 0.105 & 2019 Mar 18 & 3.0 \\
22.313354$^{3}$ & $5_{-3} \longrightarrow 6_3$~E (v$_t$=0)  &  54 & 0.105 & 2019 Mar 12 & 3.0 \\
22.365338$^{3}$ & $34_{10} \longrightarrow 35_{10}$~E (v$_t$=2 $\longrightarrow$ 1) &  54 & 0.105 & 2019 Mar 18 & 3.0 \\
22.644249$^{2}$ & $21_5 \longrightarrow 21_9$~A$^{+}$ (v$_t$=1 $\longrightarrow$ 0) &  53 & 0.103 & 2019 Mar 18 & 4.0 \\
22.756066$^{3}$ & $22_2 \longrightarrow 23_{-1}$~E (v$_t$=2) & 52 & 0.103 & 2019 Mar 18 & 4.0 \\
22.845931$^{3}$ & $36_{11} \longrightarrow 35_9$~A$^{+-}$ (v$_t$=0 $\longrightarrow$ 1)  & 52 & 0.103 & 2019 Mar 18 & 4.0 \\
22.880797$^{3}$ & $15_2 \longrightarrow 16_0$~A$^+$ (v$_t$=0) & 52 & 0.103 & 2019 Mar 18 & 3.0 \\
22.895592$^{3}$ & $10_0 \longrightarrow 9_{-3}$~E (v$_t$=1) & 52 & 0.103 & 2019 Mar 18 & 3.0 \\
23.029665$^{3}$ & $30_{19} \longrightarrow 31_{17}$~A$^{+-}$ (v$_t$=0 $\longrightarrow$ 1)  & 52 & 0.103 & 2019 Mar 18 & 4.0 \\
23.18141$^{3}$ & $31_3 \longrightarrow 31_3$~A$^{-+}$ (v$_t$=0)  & 52 & 0.101 & 2019 Mar 18 & 5.0 \\
23.200283$^{3}$ & $22_2 \longrightarrow 22_2$~A$^{+-}$ (v$_t$=0)  & 52 & 0.101 & 2019 Mar 18 & 3.0 \\
23.346879$^{3}$ & $7_1 \longrightarrow 7_1$~A$^{-+}$ (v$_t$=0)  & 52 & 0.100 & 2019 Mar 17 & 4.0 \\
23.38721$^{3}$ & $10_3 \longrightarrow 11_1$~E (v$_t$=0) & 51 & 0.100 & 2019 Mar 12 & 4.0 \\
23.779232$^{3}$ & $27_6 \longrightarrow 28_8$~E (v$_t$=1 $\longrightarrow$ 0)  & 51 & 0.099 & 2019 Mar 18 & 6.0 \\
23.83718$^{3}$ & $13_8 \longrightarrow 14_5$~A$^{+}$ (v$_t$=2) & 50 & 0.098 & 2019 Mar 18 & 7.0 \\
23.903763$^{3}$ & $29_{103} \longrightarrow 30_{101}$~E (v$_t$=2) & 50 & 0.098 & 2019 Mar 18 & 5.0 \\
23.932191$^{3}$ & $22_1 \longrightarrow 22_1$~A$^{-+}$ (v$_t$=1) & 50 & 0.098 & 2019 Mar 18 & 4.0 \\
24.013368$^{3}$ & $23_7 \longrightarrow 22_8$~A$^{+}$ (v$_t$=1) & 50 & 0.097 & 2019 Mar 18 & 6.0 \\
24.125963$^{3}$ & $34_5 \longrightarrow 33_8$~E (v$_t$=2 $\longrightarrow$ 1)  & 50 & 0.097 & 2019 Mar 18 & 7.0 \\
24.153611$^{3}$ & $30_{106} \longrightarrow 29_{107}$~E (v$_t$=1) & 50 & 0.097 & 2019 Mar 18 & 10.0 \\
24.218554$^{3}$ & $15_5 \longrightarrow 14_8$~A$^{+}$ (v$_t$=2) & 50 & 0.097 & 2019 Mar 18 & 5.0 \\
\hline
\multicolumn{6}{l}{$^1$from \citet{Lovas2004}, $^2$from the CDMS \citep{CDMS2005}, $^3$from the JPL \citep{JPL1998}, $^4$from \citet{csg05}}
\end{longtable}
\twocolumn
\bsp	
\label{lastpage}
\end{document}